\documentclass[11pt]{article}

\usepackage[margin=1in]{geometry}
\usepackage{setspace}
\usepackage{amsmath, amssymb}
\usepackage{graphicx}
\usepackage{enumitem}
\usepackage{authblk}
\usepackage{hyperref}

\title{Automated Protein Motif Localization using Concept Activation Vectors in Protein Language Model Embedding Space}

\author[1]{Ahmad Shamail}
\author[1]{Claire McWhite}
\affil[1]{University of Arizona}
\date{}

\begin{document}
\maketitle
\singlespacing

\begin{abstract}

We present an automated approach for identifying and annotating motifs and domains in protein sequences, using pretrained Protein Language Models (PLMs) and Concept Activation Vectors (CAVs), adapted from interpretability research in computer vision. We treat motifs as conceptual entities and represent them through learned CAVs in PLM embedding space by training simple linear classifiers to distinguish motif-containing from non-motif sequences. To identify motif occurrences, we extract embeddings for overlapping sequence windows and compute their inner products with motif CAVs. This scoring mechanism quantifies how strongly each sequence region expresses the motif concept and naturally detects multiple instances of the same motif within the same protein.  Using a dataset of sixty-nine well-characterized motifs with curated positive and negative examples, our method achieves over 85\% F1 Score for segments strongly expressing the concept and accurately localizes motif positions across diverse protein families. As each motif is encoded by a single vector, motif detection requires only the pretrained PLM and a lightweight dictionary of CAVs, offering a scalable, interpretable, and computationally efficient framework for automated sequence annotation.

\end{abstract}

\section*{1. Introduction}

Structural motifs and domains represent the fundamental functional units of proteins, essentially modular units that can be shuffled, duplicated, and combined across different proteins to create diverse biological functions. These conserved sequence regions, ranging from short linear motifs of 3-10 amino acids to larger domains of 50-250 residues, encode specific molecular functions from catalytic sites and binding interfaces to regulatory elements and structural scaffolds. Accurately identifying the presence or absence of motifs in proteins is essential to understand protein function. Traditional annotation pipelines including  Pfam , InterPro, and PROSITE have successfully catalogued thousands of motifs using sequence alignments, profile Hidden Markov Models (HMMs), and regular expressions \cite{Blum2024InterPro, Finn2014Pfam, Sigrist2026Prosite}. Tools such as HMMER and FIMO enable scanning for new occurrences of known motifs using profile HMMs or position-specific scoring matrices 
\cite{10.1371/journal.pcbi.1002195, 10.1093/bioinformatics/btr064}. While these approaches have been invaluable, their dependence on explicit sequence similarity and curated motif patterns limits their ability to detect conserved motifs with divergent sequences, remote homologs, or novel variants that maintain function despite sequence variation.

Protein language models offer the opportunity for fundamentally different approaches to motif detection. PLMs such as ESM-C \cite{esm2024cambrian} are trained on millions of protein sequences, and produce distributed representations that capture biochemical and structural patterns. Essentially, PLMs represent amino acids in a sequence as high-dimensional embeddings that reflect their role in the broader sequence context.  Crucially, these learned representations can reflect functional and structural similarity between proteins even when sequence identity is low, as the model captures abstract patterns rather than explicit amino acid matches \cite{Liu2024PLMSearch, Pantolini2024EBA}.

Several recent studies have explored both PLMs and other transformer models for motif and domain identification. Here, we highlight four fundamentally different approaches. MotifAE \cite{Hou2025.11.04.686576} trains sparse autoencoders on PLM representations, identifying which learned features correspond to known functional motifs. The Encyclopedia of Domains \cite{doi:10.1126/science.adq4946} determines consensus domains from the Alphafold database of protein structures, identifying more than 100 million novel domains. Chat-based models such as Evolla \cite{Zhou2025.01.05.630192} are trained to connect protein sequences directly to natural language functional descriptions, allowing users to query for protein domain presence conversationally. Finally, the InterPro-N (previously PFAM-N) model \cite{10.1093/nar/gkae997} applies a transformer architecture inspired by panoptic segmentation to assign residue-level domain labels and distinguish repeated domain instances. These approaches demonstrate the breadth of current efforts to improve motif and domain annotation, but they generally rely on specialized architectures or large-scale training pipelines. In contrast, our method takes a conceptually orthogonal perspective, focusing on how motifs can be represented directly as directions in pretrained embedding space.

Here, we introduce a novel approach in which each motif or domain is treated as a distinct \textit{concept} encoded as a direction in the PLM representation space. We adapt Concept Activation Vectors (CAVs), an interpretability framework originally developed for understanding neural networks \cite{kim2018interpretabilityfeatureattributionquantitative}, to protein sequence analysis. CAVs identify directions in the embedding space that correspond to human identifiable concepts. In the original framework, linear classifiers are trained to distinguish examples with and without a concept, such as images with and without stripes, and the vector perpendicular to the resulting decision boundary (the classifier's weight vector) serves as a CAV for that concept. The alignment between these concept vectors and a query, as measured by inner product, quantifies how strongly the query expresses the concept. For instance, images of zebras align strongly with the concept of stripes.

We apply this principle to protein motifs: by training linear classifiers to distinguish motif from non-motif sequences, we obtain CAVs that capture the concepts of protein motifs. The inner product between embeddings of sequence windows (or regions) and a motif's CAV quantifies how strongly each position expresses the corresponding biochemical function, producing alignment scores that peak at motif locations. Multiple peaks naturally indicate multiple motif instances within the same protein.

A key advantage of this approach to motif detection is its simplicity. Each motif is represented by a single vector in embedding space, and motif detection becomes a simple inner product computation between these vectors and sequence window embeddings. Notably, once a CAV is obtained for a motif, the linear classifier itself can be  discarded, as motif detection for a protein requires only an embedding produced by the pretrained PLM and one stored vector per motif. This makes the method highly interpretable, computationally efficient, and trivially extensible to new motifs.

Our objective is to demonstrate that such concept-based geometric representations can be used to localize motifs in a scalable and interpretable manner. Specifically, we show that linear classifiers trained on pretrained PLM embeddings can produce motif-specific CAVs, enabling accurate detection of known motifs directly from representation space. We evaluate this framework across a curated set of well-characterized motifs and analyze layer-wise behavior of the underlying embeddings. Our results show that this simple approach achieves ~90\% precision and ~80\% recall in motif localization, demonstrating that functional motifs are indeed encoded as interpretable linear directions in PLM representation space.

\section*{2. Methodology}

\begin{figure}[h!]
    \centering
    \includegraphics[width=1\textwidth]{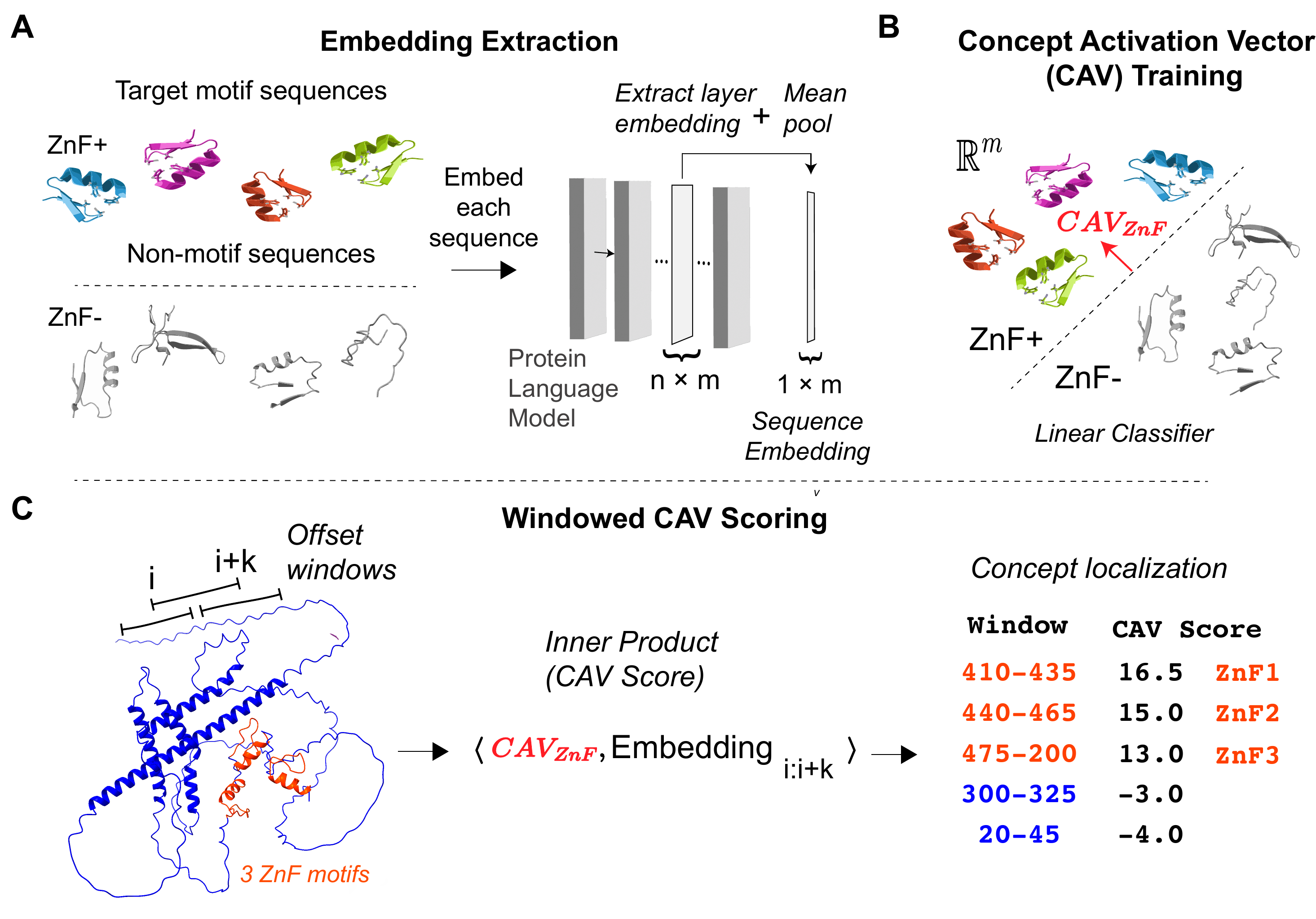}
    \caption{ Overview of the proposed motif localization pipeline. (A) We first extract pooled embeddings from protein subsequences with and without a given motif (e.g., ZnF). These embeddings are ideally taken from an mid-late layer of the PLM. (B) These positive and negative examples are then used to train a linear classifier, with the vector orthogonal to the boundary defining the Concept Activation Vector (CAV) for that motif. (C) Finally, for an unknown protein, candidate segments are aligned with the CAV via inner products, where higher scores indicate stronger evidence of the motif.
    }
    \label{pipeline}
\end{figure}

Our approach consists of two major stages: (i) learning motif-specific Concept Activation Vectors (CAVs), and (ii) applying these vectors to perform localized motif detection in unannotated sequences. Each step is designed to preserve biological interpretability while taking advantage of the representational strength of pretrained protein language models. Figure \ref{pipeline} shows an overview of the entire pipeline.

\subsection*{Stage I: Learning Motif Concept Vectors}

\paragraph{Constructing Windowed Training Samples.}
For each Pfam motif, we first construct a training dataset consisting of short subsequences (“windows”) drawn from proteins known to contain that motif (Figure \ref{pipeline}A). Positive windows are extracted by centering a window of length $w$ on each annotated motif instance, where $w$ is defined as the median annotated motif length plus a small buffer to accommodate natural variation across homologs. We also generate negative windows of identical length, sampled from proteins that do not contain the motif. All proteins and domain locations were retrieved through the UniProt API \cite{10.1093/nar/gkaf394}. This yields a balanced dataset of local contexts that do or do not contain the target motif.

\paragraph{Window embedding}

We then create embeddings for each positive and negative sequence using a Protein Language Model, in this case ESM-C \cite{esm2024cambrian} Each protein subsequence is represented as an ordered sequence of amino acids
\[
S = (a_1, a_2, \dots, a_n), \quad a_i \in \mathcal{A}, \; |\mathcal{A}| = 20,
\]
where $n$ is the subsequence length.

Passing $S$ through a pretrained protein language model yields contextualized residue embeddings.  
For a chosen model layer $m$, the activations are expressed as
\[
H_m =
\begin{bmatrix}
h_{1m} \\
h_{2m} \\
\vdots \\
h_{nm}
\end{bmatrix}
\in \mathbb{R}^{n \times d},
\]
where $h_{im} \in \mathbb{R}^{d}$ denotes the $d$-dimensional embedding of residue $a_i$.  

To obtain a fixed-size representation for each window, we must summarize its residue embeddings to a single vector. Several pooling strategies exist for this purpose, including CLS-Pooling, Mean-Pooling, and the recently proposed BoM-Pooling~\cite{Hoang2025BoMPooling}. In this work, we use Mean-Pooling, which computes the average embedding over all residues to obtain a single global representation. 

\[
\bar{h}_m = \frac{1}{n} \sum_{i=1}^{n} h_{im}.
\]

Despite its simplicity, Mean-Pooling is consistently effective in practice \cite{Vieira2025}. As motifs are local in nature, we produce an embedding for each individual subsequence, rather than embedding the whole sequence and then subsetting these embeddings to windows. Recent analyses of ESM models demonstrate that structural signals, such as residue–residue contacts, are encoded through local sequence windows, with even short stretches of sequence sufficient to recover predicted contacts \cite{doi:10.1073/pnas.2406285121}. We therefore embed  each window independently, to better capture local motif-associated signals without interference from distant regions of the protein.

\paragraph{Learning the Functional Direction.}  
Given pooled window embeddings for positive and negative samples for a target motif, we then train a linear classifier (logistic regression) to distinguish between the two distributions (Figure \ref{pipeline}B). Let the separating hyperplane be defined as
\[
w^\top x + b = 0,
\]
where $x$ is a $d$-dimensional window embedding. The normal vector $w$ orthogonal to this boundary, oriented toward the positive class, captures the direction in embedding space that best characterizes the motif.  
We interpret this vector as the motif’s \emph{Concept Activation Vector (CAV)}\cite{kim2018interpretabilityfeatureattributionquantitative}, representing the direction in $\mathbb{R}^m$ that captures how strongly the motif concept is expressed in the embedding space. 

Training is performed independently for each motif, producing a dictionary of motif vectors:
\[
\mathcal{V} = \{v_1, v_2, \dots, v_K\}, \quad v_k \in \mathbb{R}^{d},
\]

where each $v_k$ corresponds to a learned motif concept. This stage effectively converts qualitative biochemical properties (e.g., “zinc-finger-ness”) into quantitative geometric directions within the model’s representation space. Prior work has shown that neural embedding spaces can encode semantic relations as approximately linear directions, with relation-specific vector offsets supporting vector-style reasoning in the embedding space \cite{mikolov-etal-2013-linguistic}. In our framework, motifs are represented as interpretable directions in embedding space, and these motif vectors can be directly used to score new sequences, identifying regions that align strongly with a given motif concept.

\subsection*{Stage II: Motif Localization via Concept Alignment}

\paragraph{Window-Level Embedding Extraction.}  

To detect and localize functional motifs within proteins, we embed short windows of subsequences (Figure \ref{pipeline}C).

Formally, every window of length $w$ beginning at position $t$ is fed to the pretrained model as a standalone sequence,
\[
S_{t:t+w}.
\]

The window size $w$ is motif-specific: for each Pfam family, we compute the median annotated length from the training data and add a small buffer to account for natural variation across homologs. This yields motif-appropriate receptive fields without excessively enlarging the search space. We slide this window with a stride set to one-half of its length, ensuring dense coverage and allowing overlapping windows to compete for high alignment scores. Such overlap is essential, since true motif boundaries rarely align perfectly with a fixed grid and often require multiple shifted windows to capture their maximal signal.

Its internal activations $H_m^{(t)} \in \mathbb{R}^{w \times d}$ are then pooled to produce a single representation,
\[
\bar{h}_{t,m} = \frac{1}{w} \sum_{i=1}^{w} h_{t,m}^{(i)}.
\]

\paragraph{Computing Concept Alignment Scores.}  
Given the window embedding $\bar{h}_{t,m}$ and the learned CAV $v_k$ corresponding to motif $k$, we measure the degree of concept alignment through the inner product:
\[
s_{k,t} = \langle v_k, \bar{h}_{t,m} \rangle.
\]
This alignment score reflects how closely the representation of the current window expresses the functional motif concept.  Higher scores indicate stronger motif-like signal in that region. We thus rank candidate motif locations in the protein by CAV score. 

\subsection*{Code Availability }

Relevant code to perform analyses is available at \href{https://github.com/A-Shamail/TCAV/}{https://github.com/A-Shamail/TCAV/}.

\section*{3. Results}

\begin{figure}[h!]
    \centering
    \includegraphics[width=1\textwidth]{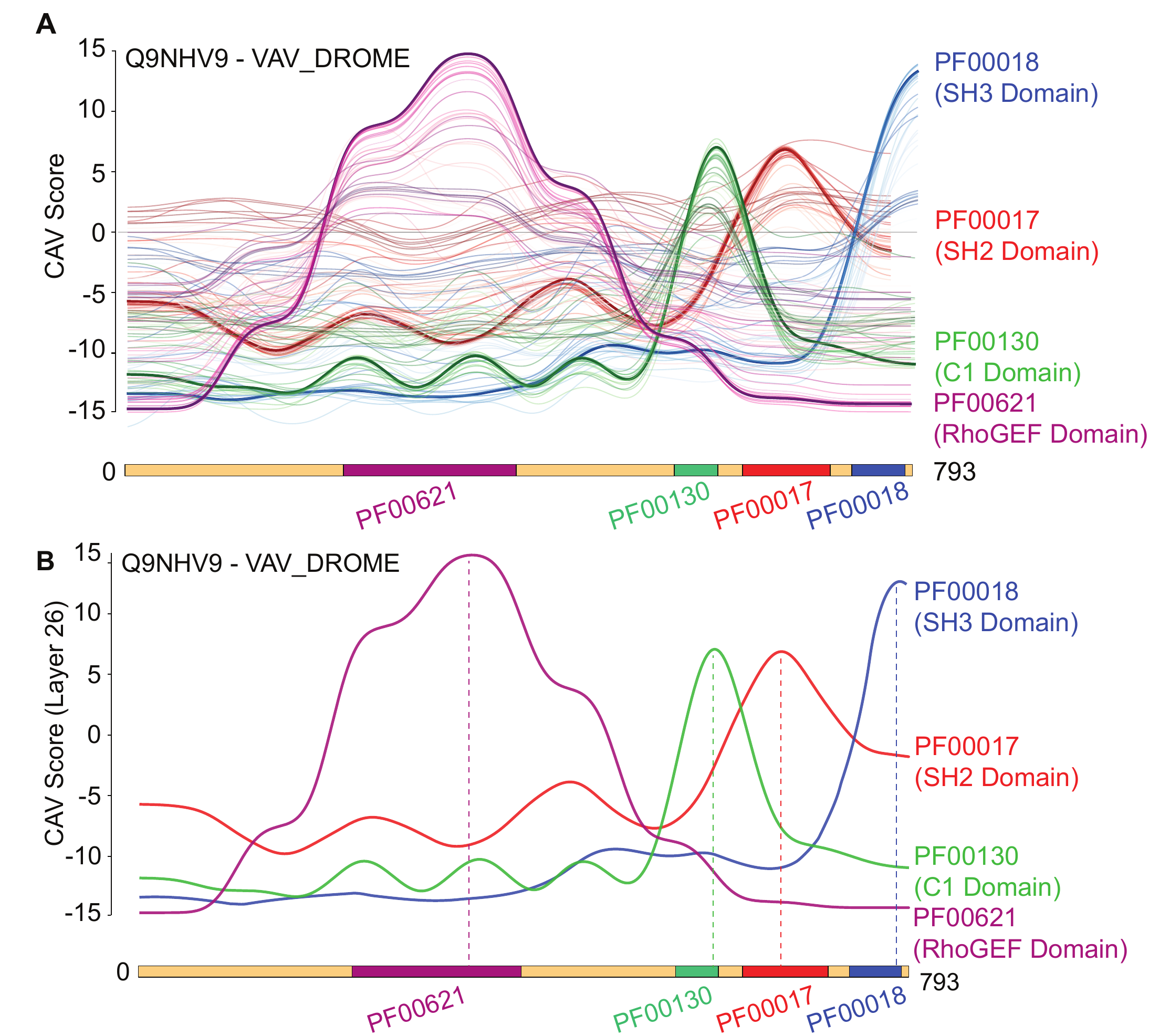}
    \caption{ (A) Layerwise CAV score profiles for all 36 layers of ESM-C 600M. Bold curves denote CAV scores for layer 26 (indexed from 1). Excluding the bold curves, earlier layers are lighter and later layers are darker. (B) CAV peaks fall in ground-truth domain intervals.}
    \label{window_peaks}
\end{figure}

\paragraph{Choice of Layer for CAV Scoring.}
We first examine how the choice of model layer affects the quality of the CAV signal. Figure~\ref{window_peaks}A shows smoothed CAV score profiles produced by all 35 layers of ESM-C on the example protein. Each amino acid is assigned the mean alignment score of all windows in which it appears, producing a continuous curve that reflects how strongly each motif is expressed at every position. Although all layers follow the same broad trend, each responding to the same underlying motif structure, the magnitude, sharpness, and signal-to-noise ratio differ substantially across the network.

A consistent pattern emerges: early layers exhibit weak motif-specific signal, mid-level layers (approximately layers 21--31) produce the strongest and most discriminative peaks, and the deepest layers (32-36) again show diminished performance. This symmetric behavior suggests that motif-related information is represented most cleanly in the middle layers, where local biochemical features and longer-range dependencies are jointly available but not yet saturated by high-level abstraction. This observation is consistent with a broad body of representation-learning research showing that middle layers of deep networks concentrate semantically meaningful structure, while early layers encode low-level patterns and late layers collapse information into task-specific summaries. This hierarchical pattern was first characterized in CNNs through feature-visualization studies \cite{zeiler2013visualizingunderstandingconvolutionalnetworks}, later generalized to deep neural representations and feature transferability \cite{yosinski2014transferablefeaturesdeepneural}.

Among these layers, layer~26 (indexed from 1, highlighted in Figure~\ref{window_peaks}A) repeatedly appears as one of the highest-scoring and most stable across motifs. For this reason, all subsequent analyses and visualizations in this work use layer~26 as the default layer for computing alignment scores.

\paragraph{Example Localizations on Multi-Domain Proteins.}
Using layer~26, we illustrate the behavior of the CAV-based localization procedure on a representative protein, Q9NHV9/VAV\_DROME, which contains four annotated domains for which concept vectors were trained.  Figure~\ref{window_peaks}B shows the smoothed CAV score profiles for each motif across the full sequence (just for layer 26).

Across all four motifs, the score trajectories exhibit the expected pattern: each motif attains a clear peak within its annotated ground-truth interval and remains low elsewhere. The transitions in the score profiles, both rising and falling, follow the boundaries of the annotated domains, indicating that the method is sensitive to motif locations rather than merely detecting coarse or diffuse signal. Qualitatively, a CAV score threshold of approximately five in this case appears to delineate motif-containing regions with high fidelity,  capturing the annotated intervals while excluding adjacent non-motif residues. 

\begin{figure}[t]
    \centering
    \includegraphics[width=1\textwidth]{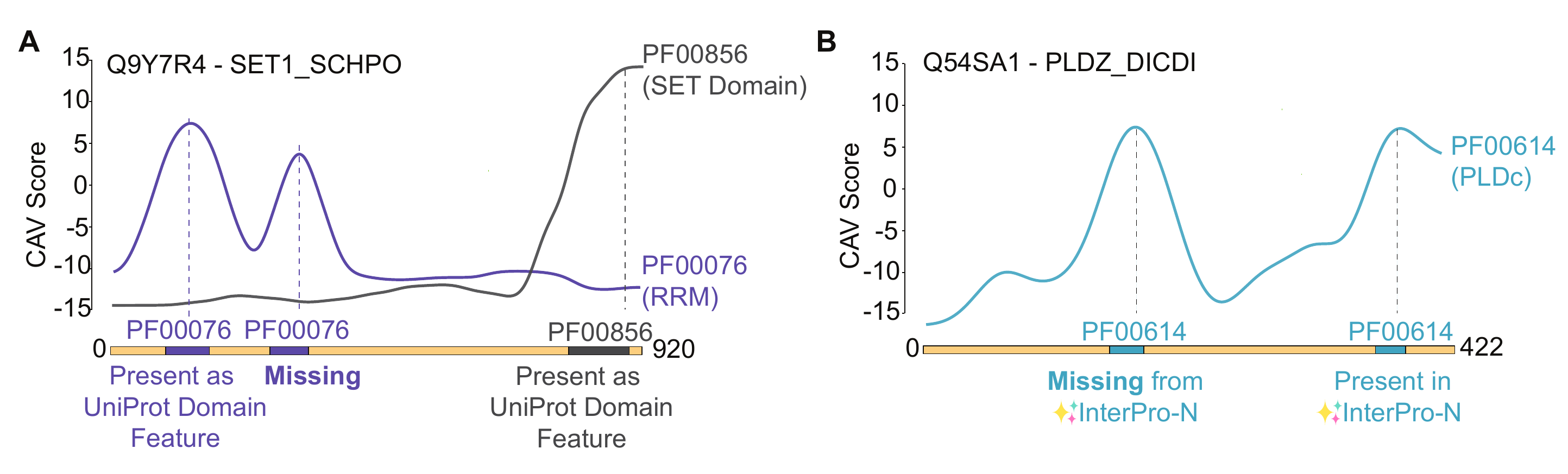}
    \caption{ (A) CAV profiles can distinguish multiple occurrences of a motif or domain in a protein (A) CAV detects a peak for a second PF00076 domain in SET1\_SCHPO, which is not annotated as a UniProt Domain Feature. (B) CAV detects a peak for a PLDc motif which is not detected by the  InterPro-N deep learning model.}
    \label{repeats}
\end{figure}

Figures \ref{repeats} demonstrates cases where the same domain or motif is present multiple times in the same proteins. These examples demonstrate that motif-specific concept vectors produce sharp, well-localized signals even in proteins with multiple occurrences of the same domain. Interestingly, the second peak of PF00076 RRM  is missing from the current UniProt Domain Features for Q9Y7R4/SET1\_SCHPO (Figure \ref{repeats}A). As we retrieved domain/motif coordinates from the UniProt API, this second peak initially appeared as a false positive in our evaluation. However, this second RRM is actually detected by several other detection methods and databases, including SMART Domains \cite{10.1093/nar/gkaf1023} and the  InterPro-N  model \cite{cheng2021perpixelclassificationneedsemantic, 10.1093/nar/gkae997}. Similarly, for Q54SA1/PLDZ\_DICDI, while the first occurrence of PF00614 is clearly detected by a CAV peak, it is missed by InterPro-N (Figure \ref{repeats}B).     

More broadly, these examples demonstrate the potential for CAVs to recover missed or weakly supported domain annotations at scale. Because they operate on residue-level embeddings rather than predefined sequence profiles, CAVs can highlight windows whose local representations resemble known motifs even when alignment-based methods offer limited signal. When applied systematically across proteomes, CAVs could offer a complementary signal to existing annotation systems and enable the continuous refinement of protein domain databases.

One consideration is that arrays of immediately adjacent occurrences of the same motif are sometimes localized by just one broader peak, with the array region generally covered by a CAV score of $>= 5$. More examples of CAV score profiles on proteins can be found in Appendix A.

\paragraph{Evaluation Across Motifs.}
To assess performance at scale, we constructed a benchmark consisting of ten randomly selected proteins per motif, yielding approximately 690 proteins in total. For each protein, we computed CAV scores for all 69 motifs and produced a ranked list of predicted windows. For evaluation, we adopted a simple but intuitive strategy: for each protein, we retained only the top-$k$ scoring windows, where $k$ equals the number of ground-truth motif instances. 

We evaluated performance under varying overlap thresholds, summarized in Table~\ref{metrics}. A predicted window is counted as a true positive if its overlap with a ground-truth motif exceeds a threshold defined as
\[
\text{Overlap} = \frac{\lvert \text{prediction} \cap \text{ground truth} \rvert}{\lvert \text{ground truth} \rvert} \times 100.
\]

The overall trend is consistent: performance remains strong at lenient to moderate thresholds, with F1-scores of 0.88, 0.87, and 0.87 at the 10\%, 20\%, and 30\% overlap thresholds, respectively. Even at a 60\% requirement, the method maintains a respectable F1-score of 0.80. Beyond this point, however, performance drops sharply, falling to 0.59 at 80\% and 0.19 at 100\%. This behavior indicates that the method very reliably identifies \emph{where} motifs occur, but is less precise at reproducing their exact annotated boundaries.

\begin{table*}[h!]
\centering
\caption{Motif detection performance across overlap thresholds.}
\label{metrics}
\begin{tabular}{|c|ccc|ccc|}
\hline
\textbf{Overlap} & \textbf{Precision} & \textbf{Recall} & \textbf{F1} & \textbf{TP} & \textbf{FP} & \textbf{FN} \\
\hline
10\% & 0.9403 & 0.8289 & 0.8811 & 867 & 55  & 179 \\ % file9
20\% & 0.9315 & 0.8317 & 0.8788 & 870 & 64  & 176 \\ % file8
30\% & 0.9231 & 0.8260 & 0.8718 & 864 & 72  & 182 \\ % file7
40\% & 0.9191 & 0.8260 & 0.8701 & 864 & 76  & 182 \\ % file6
50\% & 0.8920 & 0.8136 & 0.8510 & 851 & 103 & 195 \\ % file5
60\% & 0.8248 & 0.7830 & 0.8033 & 819 & 174 & 227 \\ % file4
70\% & 0.7197 & 0.7046 & 0.7121 & 737 & 287 & 309 \\ % file3
80\% & 0.5921 & 0.5899 & 0.5910 & 617 & 425 & 429 \\ % file2
90\% & 0.4054 & 0.4054 & 0.4054 & 424 & 622 & 622 \\ % file1
100\% & 0.1960 & 0.1960 & 0.1960 & 205 & 841 & 841 \\ % file0
\hline
\end{tabular}
\end{table*}

This limitation is not entirely surprising. Motif boundaries in curated databases are not always sharply defined, and many motifs have variable extents across homologs. Thus, the observation that predictions cluster strongly in the correct regions, yet do not always match the exact annotated endpoints, is consistent with the biological ambiguity inherent in motif definition.

Taken together, these results demonstrate that CAV-based scoring provides a reliable and biologically meaningful signal for localizing motif-rich regions across a diverse set of proteins. While the precise delineation of motif boundaries remains challenging, the method consistently identifies the correct regions with high confidence. These findings establish a strong foundation for more refined boundary-aware approaches in future work.

\section*{4. Conclusion}

In this work, we show that functional protein motifs can be represented as coherent concepts within the embedding space of a pretrained protein language model. By training simple linear classifiers on motif and non-motif subsequences and extracting their normal vectors as Concept Activation Vectors, we find that motifs align with clear, interpretable directions that produce sharp localization signals across diverse protein families. These results indicate that many conserved structural patterns are embedded in PLM representations.

This perspective offers a compelling alternative to both alignment-based annotation and more complex end-to-end neural architectures. Unlike methods requiring persistent complex architectures, our entire detection system reduces to embeddings produced by pretrained PLM and a collection of motif vectors, one per concept. Once a motif-specific CAV is learned, detection reduces to a single inner product per window, making the approach highly interpretable, computationally lightweight, and readily extensible to large motif libraries. The consistency of our domain localization results suggests that PLMs encode structural regularities in a surprisingly structured and disentangled manner, with individual motifs occupying distinct, linearly separable regions of the representation space. Taken together, this framework provides a simple, scalable route for motif annotation that leverages information already present in modern PLMs.

\newpage
\section*{Appendix A}
\begin{figure}[h!]
    \centering
    \includegraphics[width=0.92\textwidth]{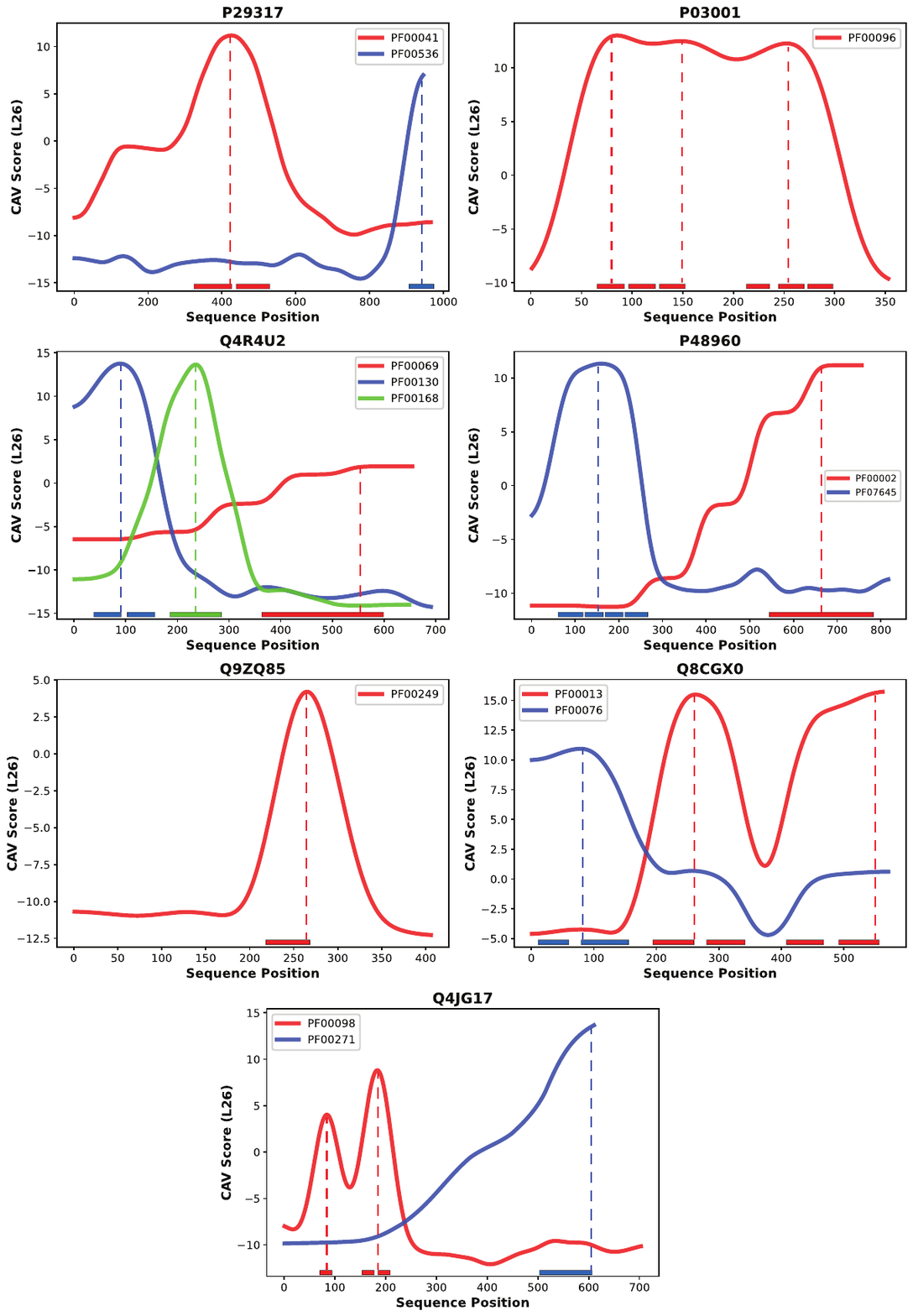}
    \label{Appendix}
\end{figure}

\newpage
\bibliographystyle{plain} 
\bibliography{references}

\end{document}